\documentclass[sigconf]{acmart}

\AtBeginDocument{%
  }

\copyrightyear{2026}
\acmYear{2026}
\setcopyright{cc}
\setcctype{by}
\acmConference[UIST Adjunct '26]{The 39th Annual ACM Symposium on User Interface Software and Technology}{November 02--05, 2026}{Detroit, MI, USA}
\acmBooktitle{The 39th Annual ACM Symposium on User Interface Software and Technology (UIST Adjunct '26), November 02--05, 2026, Detroit, MI, USA}
\acmDOI{10.1145/3830397.3841903}
\acmISBN{979-8-4007-2855-6/2026/11}

\begin{document}

\title{Lookover: Scaffolding Student Participation in Social Annotation}

\author{Yubo Zhou}
\email{yubozhou@umich.edu}
\affiliation{%
  \institution{University of Michigan}
  \city{Ann Arbor}
  \state{MI}
  \country{USA}
}

\author{Xinyi Xu}
\email{xinyixuu@umich.edu}
\affiliation{%
  \institution{University of Michigan}
  \city{Ann Arbor}
  \state{MI}
  \country{USA}
}

\author{Zesen Zhao}
\email{hymanzzs@umich.edu}
\affiliation{%
  \institution{University of Michigan}
  \city{Ann Arbor}
  \state{MI}
  \country{USA}
}

\author{Anhong Guo}
\email{anhong@umich.edu}
\affiliation{%
  \institution{University of Michigan}
  \city{Ann Arbor}
  \state{MI}
  \country{USA}
}

\author{Xinyue Chen}
\email{xinyuech@umich.edu}
\affiliation{%
  \institution{University of Michigan}
  \city{Ann Arbor}
  \state{MI}
  \country{USA}
}

\renewcommand{\shortauthors}{Zhou et al.}

\begin{abstract}
Social annotation platforms such as Perusall and Hypothesis are widely used to support student engagement with course readings, often through required commenting. Yet visible participation does not necessarily produce sustained or substantive peer exchange. Analyzing six weeks of Perusall activity from a graduate course (2,873 comments), we found that participation was often fragmented: 53.5\% of comments received no reply, 38.3\% were posted within the final 12 hours before the deadline, and over one-third of a double-coded sample went no further than repeating or agreeing with the reading or a peer. Interviews pointed to two linked difficulties: locating a discussion worth entering and contributing something that meaningfully advances it. We present Lookover, an AI-assisted annotation system that scaffolds both stages without writing for students. Discussion Overview helps students identify active themes and relevant threads, while Spark Thinking poses questions that encourage them to move beyond reaction and restatement toward questioning, extending, and substantively responding to peers. A preliminary study with five participants found Discussion Overview valuable for those who noticed it, while Spark Thinking's usefulness depended on how much a participant already knew about the topic.

\end{abstract}

\begin{teaserfigure}
\centering
  \includegraphics[width=0.65\textwidth]{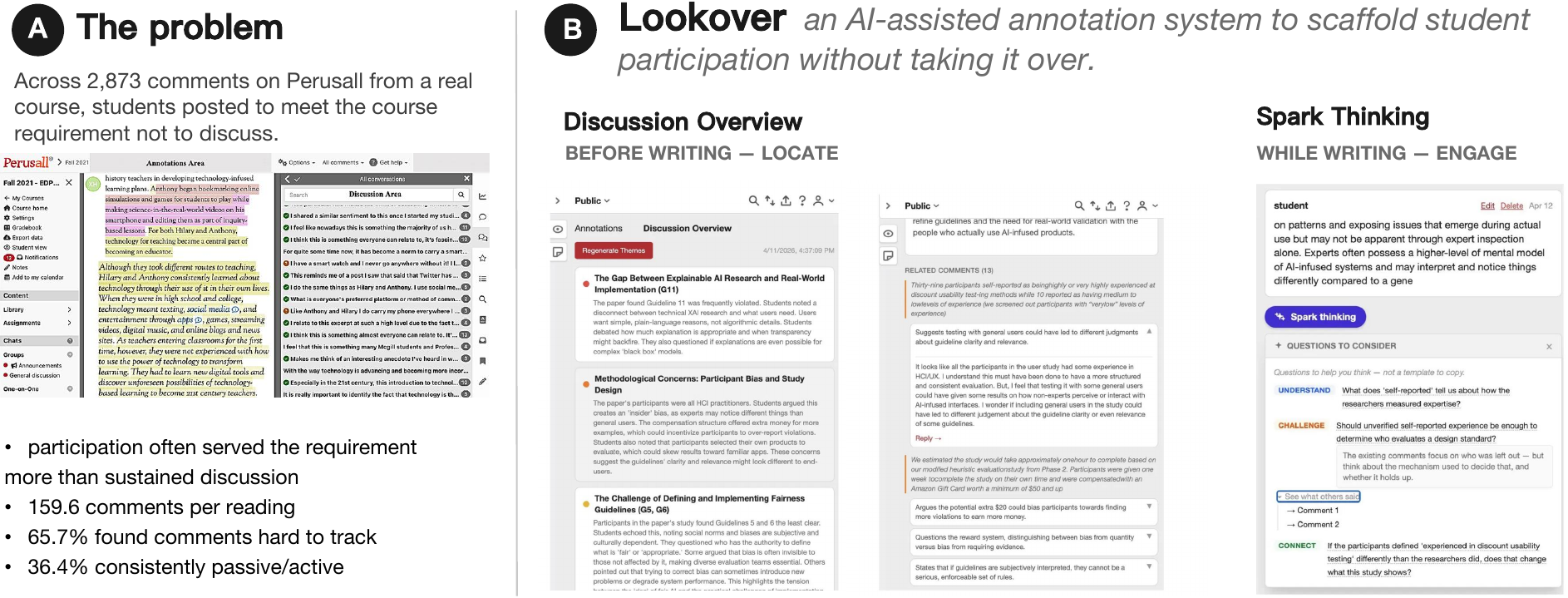}
  \caption{Lookover. Left: Discussion Overview groups comments into themes. Right: Spark Thinking poses questions when a user opens the comment editor.}
  \Description{Two screenshots of the Lookover sidebar. On the left, the Discussion Overview panel lists several themes, each with a title and a short summary of what students argued. On the right, the Spark Thinking panel shows questions at three levels---Understand, Challenge, and Connect---displayed above the comment editor.}
  \label{fig:teaser}
\end{teaserfigure}

\ccsdesc[500]{Human-centered computing~Interactive systems and tools}

\maketitle

\vspace{-3mm}

\section{Introduction}

Social annotation platforms such as Perusall and Hypothesis allow students to comment directly on course readings and respond to peers~\cite{cui2024empowering}. These activities are often structured around required numbers of annotations, which can encourage reading participation~\cite{miller2018use}. However, visible comments and participation counts do not necessarily produce sustained or substantive exchange. Prior work has proposed scaffolds such as collaboration scripts to improve response quality~\cite{deboer2021collaboration} and mechanisms to help students navigate large volumes of annotations~\cite{almahmoud2024enhancing}. Yet we know less about how participation unfolds under routine comment requirements and whether it develops into meaningful peer exchange.

To investigate this question, we collected six weeks of Perusall logs from a 55-student graduate course at an R1 university, comprising 2,873 comments across 18 readings. We adapted the ICAP framework~\cite{chi2014icap} into a codebook and two coders independently annotated a random sample of 500 comments. In the full log, 53.5\% of comments received no reply and 38.3\% were posted within the final 12 hours before the deadline; over one-third of a double-coded sample (36.4\%) went no further than restating the reading or a peer.
 A survey of 35 students echoed this: only 31.4\% agreed that discussions felt like a real exchange of ideas rather than posting in parallel. Together, these findings point to two difficulties students face: locating where a discussion is happening and where their own contribution fits, and moving beyond restating the reading to engage constructively.

We present \textbf{Lookover}, which uses AI to scaffold students into discussion rather than write it for them, addressing each in turn. \textsc{Discussion Overview} organizes existing comments into themes so students can see what peers are discussing and where to join; \textsc{Spark Thinking} poses questions as a student writes, prompting them to question and extend rather than restate, while leaving the answering to them.
\vspace{-3mm}

\section{Understanding Perusall Participation on a Real Course}
\label{sec:formative}
 We studied a graduate course on Perusall at an R1 university over six weeks, covering 55 students, 18 assigned readings, and 2{,}873 comments. The course required at least three constructive comments per reading, due before class. We combined three sources: a \textbf{descriptive analysis} of the annotation log, a \textbf{content analysis} coding individual comments for engagement quality, and \textbf{interviews and a survey} that let students reflect on their participation experience. 

\textbf{Descriptive Analysis.}
Participation stayed close to the required minimum. The average reading drew 159.6 comments, near the \textasciitilde165 expected from three comments per student, and most comments (72\%) arrived in the final 24 hours before the deadline. Comments were also largely isolated: 46\% sat on a passage where no other student commented, and peer endorsement was almost absent (only 4.1\% received any upvote, none more than two).

\textbf{Content Analysis: }We adapted the ICAP framework~\cite{chi2014icap} by grouping comments into two broader levels. Comments that primarily reacted to, restated, or directly applied the reading or a peer's point were coded as \emph{passive/active}, while comments that introduced something new or substantively built on a peer's contribution were coded as \emph{constructive/interactive}. Two coders independently annotated a random sample of 500 comments, considering each comment in relation to the reading and, where applicable, the preceding comments in its thread. The coders agreed that 36.4\% were passive/active and 54.8\% were constructive/interactive, and disagreed across this binary boundary on the remaining 8.8\% ($\kappa=.819$). 

\textbf{Interviews and Survey}.
During the contextual inquiry sessions, participants (N=3) described skimming the reading, stopping after meeting the required count of comments, and rarely returning to check for responses. One described the interaction as one-way, \textit{``I want to do the interaction with others, but I don't think there will be any people doing interaction with me''}, and another asked for a summary of \textit{``what people have discussed, so that I don't need to look at too many things.''}

The survey (\textbf{N=35}) echoed this. Replying to peers was reported as common (\textbf{65.8\%}), yet only \textbf{31.4\%} felt the discussion was a real exchange of ideas rather than parallel posting, and 65.7\% found it hard to keep track of comments across a reading. On AI, most preferred it to act only when asked (\textbf{68.6\%} vs.\ 22.9\% for AI acting on its own).

Together, these point to two unaddressed problems: comments often carried limited social meaning, restating the reading rather than building on peers; and annotations were hard to navigate as a shared space, dispersed across passages so students struggled to track the discussion.

\vspace{-3mm}

\section{Lookover}
Lookover is a sidebar add-on that uses AI to scaffold students into discussion rather than write it for them, addressing our two findings in turn.
It is built on Hypothesis, an open-source annotation platform, so it can be deployed in existing courses.

\vspace{-3mm}

\subsection{Discussion Overview}
Discussion Overview groups a document's comments into a few themes, each with a title and a short account of what students argued (Figure~\ref{fig:teaser}). Opening a theme shows a fuller summary with the peer comments behind it, each anchored to its passage and carrying a reply link, so a student can move from a theme into the specific exchange and reply there.
\vspace{-3mm}

\subsection{Spark Thinking}
Spark Thinking targets the second problem: moving students beyond restatement toward questioning and building on peers. Questioning is a well-established scaffold for this move---instructor prompts and question frames lead students to reason more and give more elaborated explanations~\cite{king1990, deboer2021collaboration}. We place AI in the role of the questioner. When a student opens the editor, \textbf{Spark Thinking} generates questions at three levels: \textbf{Understand} (what the text or a peer claims), \textbf{Challenge} (the reasoning behind it), and \textbf{Connect} (links to elsewhere in the reading), each drawn from the passage and surrounding comments, and a \textbf{See what others said} link ties it to specific peer comments the student can respond to. By design, it supplies the question but not the answer so students think and write their own response.

\vspace{-3mm}
\section{Preliminary Evaluation and Future Plan}
Five participants used Lookover on a real reading without being introduced to its features. Discussion Overview was the most valued feature among those who found it, though the participant who scrolled linearly, and would have benefited most, never discovered it. Spark Thinking's effect depended on what a participant already had to work with: one who knew the topic well found that it raised angles they had not considered, while the participant who rated it highest was unfamiliar with the topic, used it throughout, yet still produced only brief reactions.

We plan to pursue two directions. First, whether participation can extend past submission: this requires changing how the assignment is set up, not the interface, so we plan to interview instructors on why they set a fixed comment count and whether they would adopt an alternative that gives students a reason to return and respond. Second, whether better discussion means better reading: we plan to deploy the system in real classrooms and, alongside discussion quality, use an AI-free comprehension test to check whether students actually understood the reading.

\vspace{-3mm}

\begin{acks}
We thank all participants, and Muzhe Wu, Ellie Seehorn, and the research community for their constructive feedback.
\end{acks}

\bibliographystyle{ACM-Reference-Format}
\bibliography{sample-base}

\end{document}